Toward a Sensitivity-based Implicit Measure of Patients' Important Goal Pursuits

Shiva Rezvan          John A. Bargh

Albertus Magnus College     Yale University

**Abstract**

When individuals arrive to receive help from mental health providers, they do not always have well-specified and well-established goals. It is the mental health providers' responsibility to work collaboratively with patients to clarify their goals – in the therapy sessions as well as life in general -- through clinical interviews, diagnostic assessments, and thorough observations. However, recognizing individuals' important life goals is not always straightforward. Here we introduce a novel method that gauges a patient's important goal pursuits from their relative sensitivity to goal-related words. Past research has shown that a person's active goal pursuits cause them to be more sensitive to the presence of goal-related stimuli in the environment – being able to consciously report those stimuli when others cannot see them. By presenting words related to a variety of different life goal pursuits very quickly (50 msec or less), the patient would be expected to notice and be aware of words related to their strongest motivations but not the other goal-related words. These may or may not be among the goals they have identified in therapy sessions, and the ones not previously identified can be fertile grounds for further discussion and exploration in subsequent therapy sessions. Results from eight patient volunteers are described and discussed in terms of the potential utility of this supplemental personal therapy aid.

Towards a Sensitivity-based Implicit Measure of Patient Motivations

One of the most crucial elements of psychotherapy is the setting of clear, attainable goals for the therapy process, to help both clients and providers measure and monitor progress towards the desired outcomes (Prout, Wadkins, & Kufferath-Lin, 2021). Yet formulating these therapeutic goals can be challenging. Clients do not always have clear and specific goals when they arrive at therapy sessions. Sometimes, they do not even know what their goals and motives are and expect the therapists to clarify their motives and goals to them.[1] This has long been recognized in psychotherapy, which has a centuries-long tradition of developing methods such as projective tests and free association techniques to gain access to these hidden goals and motives (see Crabtree, 1993; McClelland, Koestner, & Weinberger, 1989; Perry & Laurence, 1984). The Thematic Apperception test, the Rorschach Inkblots, and Freud's free-association technique were all ways to induce the patient to reveal implicitly what they could or would not reveal to the therapist explicitly.

Gaining insight into the patient's important goals, not just for the therapy process but for life in general, is important for many reasons. Motivational science research of the past 25 years has revealed the remarkable power of an individual's currently active goal pursuits over a wide array of cognitive, emotional, and behavioral processes (Bargh, 2022; Heckhausen et al., 2021; Huang & Bargh, 2014; Kruglanski et al., 2014). The person's current goal directs what they pay attention to (prioritizing information relevant to that goal pursuit over other stimuli), and changes how objects, events, situations, and other people are evaluated. Subtle priming of one goal versus another (achievement versus socializing) changes who college students list as their best friends (Fitzsimons & Shah, 2008), and female students' behavioral intentions to use health-risky tanning salons and diet pills were significantly stronger if they'd just browsed an online dating site compared to a control condition (Hill & Durante, 2011). Verbal goal priming – in which a particular experimental goal is activated without the participant's awareness, and then drives the participant's behavior in the predicted direction is a reliable and robust effect, as concluded by a recent meta-analysis of nearly 350 studies (both published and 'file-drawer' variety; Weingarten et al., 2016). Even automatic or implicitly measured attitudes change (become more positive) towards stimuli as mundane as words beginning with the letter "C" if participants are currently playing a game in which those particular words give them extra points (Ferguson & Bargh, 2004).

---

[1] This lack of complete introspective access to the underlying reasons for one's choices and behavior is true of people in general, not just therapy patients -- see Nisbett & Wilson, 1977; Wilson, 2002, for reviews; see Nolan et al. 2008, and Bohns & Sommers, 2019, for compelling demonstrations.

As for the power of the active goal to alter one's behavior, experimentally turning on or off specific behavioral goals, such as to help another participant on an assigned task, changes a person's willingness to help another person even outside of the experimental context (Bargh, Green, & Fitzsimons, 2008, Experiment 3). And subtle reminders to actual Swiss investment bankers, on a Saturday morning at home, about their office workplace changes their 'honor system' self-report of successful coin flips for money, from honest to dishonest, and from fair-minded to greedy (Cohn, Fehr, Marechal &, 2014). Yet the transformative power of the current goal was never more dramatically demonstrated than in the classic "Good Samaritan" study of Darley and Batson (1973), in which Princeton seminary students who had been made late for their next class, and thus possessed a strong active goal to get there on time, rushed right by a sick, prone individual lying in the corridor.

One's current goals have these powers whether the person is aware, and consciously intends them (e.g., Locke & Latham, 1990), or not (Bargh, Gollwitzer, & Oettingen, 2010). In both cases, an active goal pursuit follows the same process, activates and employs the same brain regions, and guides attention, choices, and behavior in the same ways (see Cresswell et al., 2013; Marien et al., 2012; Pessiglione et al., 2007; review in Bargh, 2022). Most impressive on this point is that Locke and Latham's (1990) Goal Setting Theory, which was developed from decades of research on exclusively conscious, intentional goal pursuit, applies equally well to goals operating outside of awareness (Chen et al., 2021). The same researchers conducted a further decade of research and their meta-analysis of these studies concluded that the same predictions, involving the same key mediators and moderators, are supported when the goal is induced outside of the participant's awareness. Moreover, much of this research was conducted in field settings, and the researchers further concluded that the effect in real-life settings was significantly stronger than in laboratory studies.

Thus, it can often be the case that the patient's important goals and motives are operating outside of their awareness. This is manifested by their reports in therapy sessions, when they express confusion with their behaviors, feelings, and actions. Because they lack explicit access to the problem, they report more diffuse, often affective problems without a clear notion of what is causing them. For example, patients might report they do not know why they act the way they do. Or they might say they know what is best for them, but their actions don't correspond to these explicit best interests. These disconnects between their explicitly stated goals and their actual behavior can create challenges for their mental health providers when collaborating on specific therapeutic goals.

In such cases, there is a clear need for an access route to the problematic goals and motives. Given the powerful role of active goals over so many aspects of psychological (cognitive, evaluative, emotional) and behavioral functioning (Huang & Bargh, 2014), bringing these into the light can help set the goal of the therapy process. Prout et al. (2021) argued that most therapists move towards goal setting without a clear structure. Goal setting is often a challenging task. Although we, as mental health providers, are trained to listen to what is not shared in the session (Prout et al, 2021), and to "make the invisible visible" (Prout et al., 2021, p. 424), it is nonetheless difficult and not an exact science to decipher a person's important goals and motives when they themselves are not aware of them, or if they do not share them in therapy.

Regardless of the difficulties, goals for the therapy, a collaborative sense of what is to be accomplished, should be explicitly established early on in order for the patient to be able to feel that progress towards them is being made (and for the therapist to be able to point out such progress along the way). Progress towards a goal produces positive affect and increased motivation towards the goal (Carver & Scheier, 2002). When the patient is not able to put his or her finger on the underlying problem, methods to discover hidden goals and motives may be a boon to the therapeutic process.

Of course, the patient is often aware of his or her problem areas and has clear and explicit goals for the therapy process, knowing and able to express what they hope to accomplish from the sessions. Although they are aware of what the problem is, they are not able at present to deal with it effectively, perhaps because they are using the wrong approach. For example, a father wants his son to respect him and so asserts his authority and assumed power in the role of father to gain this respect, but this only seems to make matters worse. Or a patient wants expressions of love from a spouse or family member but shows this neediness to an extreme that it drives the others away. There could be self-efficacy issues as well (Bandura, 1977); cases in which the patient has the right approach, knows what needs to be done but does not feel capable of doing it. In such cases a 'life coaching' approach is called for, and the therapist can deal directly with the patient's important goals and help them find success.

Even when the patient is aware of their problematic goals or needs, they may not wish to share this information with the therapist, or they may share it selectively and censor portions of it. Of course, their desire to protect themselves emotionally is understandable, especially early on when trust has not yet developed in the therapeutic relationship. It could be a traumatic event or period in their lives that they are not ready to directly confront, or it may be a socially sanctioned interest or identity that they have kept hidden from even their close friends and family their entire life (McKenna & Bargh, 1998). Because it is not easy to make oneself vulnerable and share very personal goals and motivations at the onset of therapy, when the therapist is still perceived as a stranger, the patient may not fully disclose their important goals and motives, another reason why an indirect, implicit assessment device is useful.

Another interesting case where a window into important goals the patient is not aware of pursuing can be useful, is the case of goal conflict. A person may want to stop drinking, for example, but at the same time want to be with friends and socializing (have the important goal of socializing), and either be able to drink along with them, or feel that they require the social lubricant of alcohol to have a good time (and be a fun companion to the others). Or perhaps they engage in other health-risky behaviors, and correctly identify these as the problems they want to fix, but do not appreciate the underlying motives that are being satisfied by those risky behaviors. Kopetz and Orehek (2015) suggested that people do risky behaviors to accomplish an important accessible current goal. For example, destructive behaviors such as smoking that seem inconsistent with beneficial goals such as health may be a way for people to meet an important current goal such as conformity with peers to get accepted (Kopetz and Orehek, 2015). And so, despite agreeing to pursue the therapeutic goals, they might not take the often-difficult steps needed to attain those goals (which in the case of overcoming addictions are already hard enough to accomplish).

So, we can see that patients are often unaware of the problem areas in their lives, just of the consequences (e.g., social isolation) and symptoms (emotional disorders), and even when the patient is aware of the problem areas in their lives, they may not be aware of the important goals and motives that are contributing to these problems. In both cases it becomes important to objectively assess and access their actual current important goals and motives. Accordingly, here we report a novel technique for uncovering a patient's important but hidden goals. This effort is firmly in the long tradition of projective tests and other therapeutic tools to bring unconscious mental processes to the light of awareness; it is novel in that it leverages experimental findings over the past 20 years on the mechanics of goal pursuit and in particular its effect on attentional selectivity and perceptual sensitivity.

In this regard its origins can be traced to Bruner's (1957) classic paper on perceptual readiness, which grew out of his "New Look" research with colleague Leo Postman in the 1940s and 50s. The basic idea is that there are' top down' influences on what we see and hear, based on our current goals and purposes. Need states in particular guide our attention to those objects, people, and situations in which those needs can be satisfied. The principle of perceptual readiness led to the more specific notion of construct accessibility, in which a mental concept is more likely to be used to capture and interpret informational input the more frequently (chronic) and recently (temporary) it has been used in the past (Bargh et al., 1986; Higgins, 1996). Tversky and Kahneman's (1973) work on the *availability heuristic* is highly similar to the construct accessibility idea, as it focuses on what relevant material in memory most easily comes to mind (i.e., is most accessible) when making a judgment. Bruner's (1957) idea of perceptual readiness was more explicitly motivational in nature, however, and thus is more directly related to our present focus on discovering a patient's hidden motives.

More recently, research on the interface between unconscious and conscious mental processes (Sklar, Kardosh, & Hassin, 2021) has shown that there are three factors that determine whether unconscious contents become conscious: stimulus strength (intensity and duration), motivational relevance, and cognitive accessibility. Stimulus strength is the 'bottom-up' factor, and cognitive accessibility and goal relevance were the two types of top-down influences discussed above. In the framework proposed by Sklar et al. (2021), mental processes are inherently unconscious in nature, and are prioritized for entry into conscious awareness (which has a limited capacity or bandwidth) based on those three factors.   In their research using the 'continuous flash suppression' technique, which enables repeated subliminal stimulus presentations until the participant can consciously see and report the stimulus, the researchers showed that these three factors influence how quickly or easily a given stimulus can emerge from unconscious into conscious processing. In other words, a participant would more quickly become aware of unconsciously presented information that was relevant to their current goals and needs, compared to other information.

Goal research in the field of social cognition also shows that when a person has an active goal or need state, they become more attuned to the presence of information relevant to that goal or need. This is most dramatically shown by studies in which stimuli are presented so briefly that most people cannot see them – that is, subliminally. But if the person has an active goal for which those stimuli are relevant, then they are more sensitive to the presence of those stimuli, and they can often see or otherwise be affected by those stimuli. For example, Xu, Schwarz, and Wyer (2015; see also Aarts et al., 2001; Ferguson, 2008, for similar findings) showed that their participants were not aware of and could not report words related to eating and food presented at 50 milliseconds on a computer screen. However, if participants had not eaten for the past 4 hours, they could see those words. What was subliminal became supraliminal because of the greater sensitivity caused by the need state. In related findings on subliminal advertising, though previous experimental research had seemingly debunked claims that it could be effective in changing consumption or behavior (e.g., Greenwald, 1992), more recent research consistently shows that it can be effective when the participant has a need state corresponding to the subliminal stimuli (Strahan, et al., 2002; Karremans, et al., 2006).

Our proposed implicit measure of a patient's unconsciously operating goals and motives leverages this principle of goal pursuit in a kind of 'reverse engineering' manner. In this method, a series of words is presented that relate to various possible life goals the patient might have. On each trial, a word is flashed for a brief 50 millisecond moment, and the patient attempts to guess the word.   There is a pause of several seconds before the next word is presented, and so on. The therapist records each guess on each trial. Often the patient cannot tell what the word was, and either guesses or makes no guess, but is asked to report any of the words that they do see. Following the logic of the experimental work on goal pursuit, we would expect the patients to have greater success (hit rate) in seeing those words related to their current important goals and motives, compared to words that are not related to them. Therefore, what the patient can notice consciously is the basis for inferring backwards to what their important active goals are. These may or may not be goals that they are aware of or have discussed with their therapist. If they are, then a renewed focus and emphasis on them might be warranted. If they are not, new areas of discussion and treatment may be illuminated.

In the specific, initial version of this task we have developed, there are a total of 40 words presented. These words are synonyms of the various life task goals identified by the senior author from therapy sessions with her patients. Dozens of potential stimuli were gleaned from these sessions, which we then reduced to a set of 7 basic life-goal pursuits. We also added a group of neutral, non-goal-related words for comparison purposes.  Close synonyms of these goals were then used in the word presentation task, five words for each of the seven goal categories, and five words in the neutral group, for a total of 40 word presentation trials**.** Three of the neutral words were shown at the start of the task so that the patients could get used to the procedure. The researcher wrote down each of the patient's guesses and made further notes as to whether the patient expressed confidence in their answer, versus saying they had no idea and were just guessing.

**Method**

The method had two key components: the generation of the goal-related stimuli, and the presentation of each stimulus word to the patient-participant for a specific brief duration. Both components are modifiable to best suit the therapist's needs and to fit his or her own experience with patients. The list of stimulus words and the exact presentation duration of the word stimuli can be adjusted to adapt to differences in computer equipment (e.g., graphics capabilities) which can create variations in the actual duration of the stimulus on the monitor screen.  The monitor characteristics themselves can also impact the visibility of the stimuli – for example, the display brightness, contrast, font size, font clarity, distance of monitor from patient, the amount of ambient light in the room, and so on.

**Generation of goal-relevant stimuli**

Based on 18 years of therapy sessions with a variety of clients, a lengthy list of important goals in the patients' lives was generated. These were both goals that the patients had expressed as important to them, as well as those that the patient had inferred or suspected based on the therapeutic discussions. Extensive discussions between the two authors distilled this long list down into seven basic life-goal categories: *safety, acceptance, belonging, power, growth, existential*, and *feeling better*.  Because the specific definition of each goal is different for each patient, a broad definition of the above goals is provided here based on what most patients consciously or unconsciously shared in therapy over the years in terms of what they want to reach.

*Safety* is feeling emotionally safe while being around another person and being able to talk about life events, behaviors, thoughts. In other words, individuals seek a trusted, warm, and reliable person who gives them secure feelings. *Acceptance* refers to accepting self and others, respecting who a person is without judgment despite thoughts, feelings, and behaviors and accepting life events. *Belonging* is having somebody there for them, caring for them, listening to them, like them, or loving them. *Power* is a desire to control situations or people, being right, winning an argument, and being above other people. *Achievement* is accomplishing, producing, and being advanced and prosperous. *Existential* goals are finding meaning in life or questioning life in terms of purpose, aim, and plan. *Feeling better* is attaining a positive change in feelings, improvement in symptoms, emotions, or progress.

We should emphasize that this particular set of life goals is distilled from one therapist's (first author) long experience with patients, and so this list will likely be useful for that therapist's current as well as future patients.  Of course, this particular set of patients is not in any way sampled to be representative of all therapy patients, as they come from a certain geographic region, socio-economic status, race, gender, and so on.  These particular life and self-improvement goals are not suggested to apply equally well to any other set of patients. Certainly, they meet the 'eye test' of being basic life issues that are likely to have wide relevance and applicability, but patients' life circumstances can vary widely, and so best practice would be to adapt the set above to one's own patients' set of important life goals, following the same generative procedure we did. Alternatively, of course, the set of goal stimuli could be generated *a priori* based on one's preferred theoretical approach to therapy.

SENSITIVITY AS IMPLICIT INDICATOR OF GOALS                                                    14For each of the 8 categories, we generated 5 exemplars, for a total of 40 stimuli. Because the main outcome variable of the task is ability to correctly identify the briefly presented stimuli, we attempted to control for factors known to influence the detectability of 'subliminal' level stimuli. For example, the more frequent the word and the shorter the word, the more easily detected (lower recognition threshold) it is in general. Thus, we attempted to equate the stimuli on frequency and length across the 7 goal-related categories as much as possible. The *Corpus of Contemporary American English (COCA)*, an online tool for word frequency, was consulted for this purpose.[2] Finally, we created a random ordering of the 35 goal-related words and two control words, which were preceded by the 3 control words.

**Presentation of stimuli**

A computerized-research software program was developed to present the stimulus words, using the AMP programming code.[3] The primary languages used in AMP are JavaScript, PHP, and HTML/CSS. The program's first step was to ask for the stimulus words in order of presentation, and the researcher typed those in one at a time. The second step was to enter the duration of the word presentation on each trial. We used 50 msec based on prior research (e.g., Xu, Schwarz, & Wyer, 2015, Study 1), but because of the variability in subjective duration caused by computer equipment and monitors, this was made an input option for the therapist. The final step was to select whether the stimuli would to be pattern-masked or not, again for the same reasons of variable visibility across computer equipment. Pattern-masking reduces the duration of the iconic storage of the item, which persists for a while after presentation if not masked (see Sperling, 1960; Neisser, 1967). Pattern-masking helps ensure that the actual subjective duration of the stimulus is the same as the presentation duration. The duration of the mask was 100ms.

---

[2] https://www.english-corpora.org/coca/
[3] For more information you can visit https://dictionary.apa.org/affect-misattribution-procedure and https://compass.onlinelibrary.wiley.com/doi/pdf/10.1111/spc3.12148

**Participants**

Eight individuals currently receiving in-person psychotherapy were approached for possible participation in the word presentation task, which was described as the development of a measure that might facilitate the psychotherapy process. Seven of them showed interest and volunteered to participate. There was one more additional participant who did the test run and showed interest to participate. It was emphasized to all that participation was entirely voluntary, choosing to participate or not would have no impact on the therapy process, and that they could choose to stop the task at any time. The ages of the participants varied from 20 to 70; there were 2 males and 6 females; their diversity in terms of race and ethnicity was limited: all were White and there were no African American, Latinx, or members of other groups that are generally underrepresented in psychological research studies. [We note that this is not a limitation of the proposed measure itself, because the same method and procedure would be expected to be just as fruitful in the case of URM patients, although certainly the content of the goal-related stimuli (i.e., the patients' important life tasks and goals) could vary from those of the present sample of participants.]

**Procedure**

The participants were greeted by the researcher (first author) and asked to read and complete a consent form if they chose to participate. In the consent form, they were informed that the study's goal was to develop a new measurement technique for facilitating the process of therapy. Additional information was given about the procedure. They were informed that there would be no foreseeable risks in participating beyond those they may experience in day-to-day life, and indeed there may be benefits for their therapy. As mentioned above, it was emphasized in the consent form that their participation would be entirely voluntary and that if they refused or stopped early, the therapy sessions would be unaffected. They were also assured that their identity would remain confidential and that the consent form and the data will be kept in a confidential area in a locked cabinet.

After they had given written consent, they were seated in front of a laptop, 24 inches away from the screen. The following instructions appeared on the computer screen:

"You will be presented with random words, presented very quickly, one at a time. Say any of the words that you can see as soon as you see them. If you cannot identify the words, make a guess or say that you don't know. This will not take more than 10 minutes. Thank you in advance!"

Forty words were presented on the computer screen. Three of the control words were presented first. The other two control words were dispersed randomly among the goal-related stimuli. Following the options selected beforehand, each word was presented for 50 ms, followed by a pattern mask for 100ms. When the participants saw or guessed a word, they would say it out loud. The experimenter recorded the word guessed (or no word) on each trial. [To further protect confidentiality the patient's name did not appear on the same page as the recorded guesses, but an identification number linking the patient to those guesses was used instead, with the information linking the patients to those numbers stored in a secured location.]

At the outset (during the first few neutral word trials), most (6 of 8) participants were a bit discouraged because they realized they could not see the words easily. The researcher found it useful to give an encouragement at this point to reassure the participant that they were doing fine. After all word trials had been presented, the participants were debriefed about their experience in the study. During debriefing, they were again reassured that the program was designed to present the words so quickly to make them hard to see, so it was understandable if they could not see most of the words. It was also explained that the task was intended to shed light on their important life goals, the idea being that perhaps the words they could more easily see, among all the rest, signified important goals in their life.

Following this idea up, participants were then asked to provide any personal meanings of the words that they did see correctly. As a next step in the therapeutic process, patients were asked if they would like to discuss these particular goals with the therapist. Specifically, the therapist asked if the patient would be willing to consider using the debriefing discussion to create new conscious goals in the process of their ongoing therapy. They also were asked about their guesses, to elaborate on them to help expand and facilitate treatment towards their important current life goals.

## Results

**Participants' responses and reactions to the task**

*Participant 1* saw 24 words correctly, 2 words incorrectly, and missed the rest of the words. When the word "power" was presented, the participant reported it and said with excitement, "it's clear." This participant also reported the word *dominance* with less certainty (e.g., took off the glasses and hesitantly said the word). This participant reported each of the words "acceptance" and "safety" twice, although they were only presented once. The first time the participant said "acceptance," it was correct. When they said "acceptance" the second time, the actual word was "agreeable." The word "safety" was right the first time, and the second time the real word was "secure." (It has been long known in attention research that seeing or hearing a word in a presented sequence makes it more likely to be guessed in later trials, correctly or incorrectly; Treisman, 1959).

After the task was completed, the participant noted that no word was more evident than "power" and asked why the word "power" was the easiest to see. They were encouraged to explain the personal definition of the word "power." The person noted, "I do not mean dominance" "Power is the fullest of capabilities, manifesting the best of you; it's not domination, the fullest I can do." It also means, "Is it good enough?" Then, the patient added that power might lead to domination. After more exploration, this person concluded that waiting for the best to happen and waiting to express the fullest self were not realistic goals. After more self-reflection, this participant noted power is "here and now" and is the "natural ability." We note that for this patient, there was no previous conscious or reported goal of "power" reported during therapy. A plan to create a conversation about the personal meaning of power was created. During further discussions about other goals in later therapy sessions, this participant was highly motivated to talk about "acceptance" when it was brought up. They reported they have had difficulty accepting things and would like to set up a goal for acceptance. The goal of acceptance was never shared in the previous therapy sessions. "Safety" is another area for future exploration.

*Participant 2* saw the word "towel" that was incorrect. The actual word was "power." Because of partial recognition of the "*owe*" middle section of the word, the fragment was misread as 'towel'. If anything, this suggests that power, as a goal, was not high on this patients' set of important goals, as *power* did not come to mind even when its central fragment was correctly noticed. This participant also reported the words "liked" and "together" that had not come up previously in therapy as personal goals.

After the task, the personal meaning of the words "liked" and "together" were explored. The participant shared expectations of a good relationship, such as respect, politeness. After explorations, the participant expressed interest in looking for possibilities to accept people and become more independent of other people's behavioral patterns.

*Participant 3* saw the words "trouble" and "progress." The word "trouble" was not in the list of the word stimuli. The word presented on that trial was "approval." This participant did not see or report the rest of the words and mainly reported the letters of the pattern mask (e.g., "Q at the end," or "Q," or " Z & O"). Other times, they reported "no word" or just remained silent.

According to their reports in therapy, this participant had already accomplished a lot in life. *"Progress"* had been among the previously reported goals in the therapy sessions.

In post-task discussions, this participant elaborated on the words "trouble" and "progress," and shared feelings and actions that led to a realization that they may have a goal to look for things that go wrong. As a result, the participant would try to fix things without asking. There was an additional conversation about the word "progress." After exploration, the person realized that the unconscious message might have been "I am not enough." Progress meant to do more because "you are never enough." The goal of "progress" was re-set up in the plans for future therapy sessions, with a new realization that "we are enough."

*Participant 4* saw words incompatible with reported goals, including "power," "force," and "dominant." The participant had never reported these goals in prior therapy sessions. Most of their reported goals were related to being in peace with demands on them, and a sense of freedom. Unfortunately, it was not possible to follow-up on these guesses with this participant because of unexpected pressing crises and life stressors that needed immediate attention in that session and the following sessions.

*Participant 5* was aware of 37 words presented at 50 ms with ease and reported all the 37 words correctly. This participant missed only the words "produce," "advance," and "safety." At the end of the process, the participant provided informative feedback about how the words were perceived. For example, this participant noted they did not see whole words, and most of the time, their guesses were made based on the first and often the last letter of the words.

Because this participant was generally aware of the word stimuli and was able to correctly report all but three words, it seemed in this case that a lower (faster) presentation duration (e.g., 40, 35, or 30 ms) should have been used in their case. Discussion focused on the two of the words they were not able to see. This participant had already reported the "need" for "advance" and "produce" in therapy sessions. After discussions, this participant noted that they realized, "before I know, I set up a goal for failure." They acknowledged it was possible they lacked motivation to 'produce' and 'advance'. In the following therapy sessions, Goal conflicts, and lack of goal readiness were observed as potential factors for the lack of motivation to 'produce' and 'advance'.

*Participant 6* also saw most of the words -- all but "reliable," "plan," and "road." When reporting the word "love," their voice became shaky and emotional. "Love" was not a previously reported goal or concern in the therapy sessions. They acknowledged that the missing goals were not among their strong motivation, although they "needed" them in their life. Based on reports, conflict of goals is a possibility. A discussion about "love" is planned for future therapy sessions if their life circumstances allow for future sessions.

*Participant 7* saw 30 of the 40 words correctly. They reported the word "recognized" twice. The first report of the word "recognized" was correct, but the second time, the actual word was "prospect." After completing the test, the participant reported they remembered the words "improve" and "progress" clearly and that the measure made them feel "empowered." They also were thankful for the encouragement during the task because they thought they were "making up" the words. They shared the meaning of "improve" and "progress" in three different areas,

SENSITIVITY AS IMPLICIT INDICATOR OF GOALS                                                                22including family, work, and self-related improvements. These goals will be explored. Also, more information will be obtained about the word "recognized."

*Participant 8* was not a patient, but volunteered to test run. They wanted to help and were interested in seeing how the measure might help them in their daily lives. They reported the word "speak" twice, incorrectly as it was never presented. The actual words on these trials were "secure" and "stable." This participant also said the words "accomplish" and "accepted."

The participant later denied that they have any personal goal to "speak," though they had previously expressed uncomfortable feelings being around many people. After conversations about the meaning of the word "speak," the participant realized an important desire to speak during gatherings. This unrecognized goal had apparently made this participant uncomfortable being around people. After some discussion, the participant concluded that there was really no need to talk "all the time" or "most of the time" while being around a group of people and seemed more at ease after this realization.

**Patients' subjective experience during the task**

Patients generally found the experience positive; all said that in one form or another that it was 'fun,' 'interesting', and "empowering," and several noted that 'it makes you focus' or 'concentrate', especially after missing a word on a given trial. The one participant who was less positive (#5) had been able to see most of the words and suggested that the word presentation duration had been too long, and that this part of the procedure could be improved.

Table 1 provides information on the number of words seen and not seen and their correspondence with the previously stated or not stated goals. For example, Participant 1 saw nine words corresponding to the previously stated important goals. This participant also saw 14 words corresponding to other goals and never stated in therapy. Participant 2 saw one word that they had reported in treatment and saw seven words that were never shared as important goals. Participant 3 saw one word that corresponded to the previously stated important goals. This participant did not see any other words they previously did not report as goals. Participant 4 saw four words that aligned with previously stated important goals and saw 11 words that were never shared in therapy as essential goals. Participant 5 saw five words previously reported as therapy goals and 27 words that were never said as goals. Participant 6 saw five words they had reported as therapy goals and saw 28 words that were never reported as goals. Participant 7 saw four words corresponding to previously stated important goals and saw 23 words that corresponded to other goals and were never stated in therapy before.

Table 1 also illustrates the number of goals the participants mentioned in therapy sessions featured in the measure and had synonyms presented. For example, participant 1 reported two goals in therapy sessions with synonyms presented in the measure. This number was one for participant 2, one for participant 3, two for participant 4, two for participant 5, two for participant 6, and two for participant 7. Additional information regarding the number of words that were not seen and corresponded or did not correspond to previously stated goals are presented in Table 1.

**Follow-up sessions of the task**

Participants 5 and 8 showed interest in repeating their word-guessing task, for several reasons, including empowerment, fun, or being able to talk more about the words and associated goals. The procedure was slightly different in their second run because of the need to find a lower perceptual threshold for each participant; their prior exposure to the list of words made these more accessible and thus lowered their threshold for them. We started with the 50 msec duration (masked); if the participants saw more than six words, the task was restarted with the duration decreased to 40 or 30 msec. Although the participants saw fewer words in the second run with the briefer presentations, for the most part they saw words from the same goal-categories as before.

**Discussion**

*Main hypothesis.* Although this is too early to conclude about the function of the measure, it seems the measure is likely helpful in detecting participants' strong motivations.

*Benefits to therapy sessions.* The therapy sessions are ongoing with new realizations. I could not get to these points with a few patients without talking about the patients' goals that were detected through the measure. Applying this measure in this sample made it possible to explore important unreported goals and their meanings. During exploration, skewed goals were also detected. Feist et al. (2018) indicated that patients have difficulty solving life issues due to distorted goals, and therapists have the responsibility to help with the faulty nature of patients' goals. The measure also helped distinguish the conflict among the goals and create pathways to address them. Overall, it seems that a solid focus on patients' conscious goals or self-reported measures might not be enough. Adding other tools such as this measure that facilitate access to some unconscious strong goals and motivations might create a more productive and comprehensive approach in therapy to help patients.

*Meaning of incorrect guesses.* It is also likely that seeing or not seeing the word stimuli or even making wrong guesses might provide important information during therapy. The wrong guesses might be relevant to an individual's top-down processing.

*Meaning of not seeing words related to their stated important goals.* It is also likely that not seeing the word stimuli might provide important information during therapy. Missing words raise the possibility of a lower or lack of motivation to pursue the goal related to the words. The missing words might help a therapist explore if there are missing goals in individuals' lives.

**Future directions**

Because of individual variability in the perceptual or conscious threshold, we recommend adding a first set of 10 neutral or control words, presented initially for 50 msec, and then adjusting this threshold downwards for those participants who easily see the word stimuli at these durations. This first set of neutral words should be equated on length and frequency to the goal-related words that will follow. Also, to better understand the importance of the goals to the participants, after administrating the measure, we can ask them what words they remembered. This is because the amount of attention to a word also predicts whether it is later remembered or not. Thus, the better the memory the more attention they had paid to it and so the most important this goal is to them.

*Other possible uses of contemporary goal research: Their influence on evaluations and preferences.* The established power of the active goals and motives is not limited to attentional selectivity, which is the focus of the present manuscript. The active goal also changes likes and dislikes, and can cause positive evaluations of typically negative stimuli, and vice versa. These likes and dislikes can dramatically change as a function of the participant's current need and goal state. For example, Chassin et al. (2001) show that implicit attitudes towards cigarette and smoking related stimuli changes as a function of the current nicotine need state; participant smokers who had just satisfied their need showed negative implicit attitudes towards smoking, while those who had not smoked in a while and so were in the need state showed positive implicit attitudes towards the same set of smoking-related stimuli. Ferguson in many studies (e.g., 2008; Ferguson & Bargh, 2004; see also Melnikoff & Bailey, 2018) found that implicit attitudes towards otherwise neutral stimuli such as letter C change as a function of helpfulness/hindering of current goal. Also very relevant is Nock and Banaji's (2007) demonstration that implicitly measured (using the Implicit Association Test) attitudes towards suicide predicted actual attempts by hospitalized depressed patients, whereas explicit, self-report attitude measures did not predict these attempts. Recently, a meta-analysis of discriminative and future application of the death version of the implicit association test (D-IAT) suggested that the D-IAT may have a supplementary role in suicide risk assessment. However, a comprehensive evaluation of acute suicide risk is crucial and should not only focus on D-IAT performance (Sohn, et al., 2021).

There are several relevant experimental paradigms to examine automatic or implicit evaluations that could be employed - - for example, a speeded good/bad task (press Good or Bad as fast as possible) where the positive or negative reactions are examined by the therapist to see if interesting insights could be gained, what the patient has a good impulse towards that most people would consider bad, and vice versa.

**Conclusions**

We propose the goal-sensitivity measure as an additional tool in the therapist's toolbox, one that can supplement the patient's own conscious reports in useful ways. Those explicit self-reports cannot always provide us with access to unconsciously operating motives and goals, limiting the therapist's assessment and thus ability to set up appropriate and thorough therapeutic goals and treatment plans. Those unconsciously operating motives and goals might never come up in therapy sessions otherwise, increasing the cost and expense of the therapy and decreasing its chances of success.

SENSITIVITY AS IMPLICIT INDICATOR OF GOALS                                                                28**References**

Aarts, H., Dijksterhuis, A., & De Vries, P. (2001). On the psychology of drinking: Being thirsty and perceptually ready. *British Journal of Psychology*, *92*(4), 631-642. https://doi.org/10.1348/000712601162383

Bandura, A. (1977). Self-efficacy: toward a unifying theory of behavioral change. *Psychological review*, *84*(2), 191.

Bargh, J. A. (2022). The hidden life of the consumer mind. *Consumer Psychology Review*, *5*(1), 3-18.

Bargh, J. A., Gollwitzer, P. M., & Oettingen, G. (2010). In S. Fiske, DT Gilbert, & G. Lindzay. *Handbook of social psychology*, 268-316.

Bargh, J. A., Green, M., & Fitzsimons, G. (2008). The selfish goal: Unintended consequences of intended goal pursuits. *Social Cognition*, *26*(5), 534-554.

Bargh, J. A., Bond, R. N., Lombardi, W. J., & Tota, M. E. (1986). The additive nature of chronic and temporary sources of construct accessibility. *Journal of Personality and Social Psychology*, *50*(5), 869.

Bohns, V., & Sommers, R. (2019, July). Underestimating the Difficulty of Denying Someone Access to Sensitive Data. In *Academy of Management Proceedings* (Vol. 2019, No. 1, p. 11771). Briarcliff Manor, NY 10510: Academy of Management.

Bruner, J. S. (1957). On perceptual readiness. *Psychological review*, *64*(2), 123.

SENSITIVITY AS IMPLICIT INDICATOR OF GOALS                                                         29

SENSITIVITY AS IMPLICIT INDICATOR OF GOALS                                                                31Huang, J. Y., & Bargh, J. A. (2014). The selfish goal: Autonomously operating motivational structures as the proximate cause of human judgment and behavior. *Behavioral and Brain Sciences*, *37*(2), 121.

Karremans, J. C., Stroebe, W., & Claus, J. (2006). Beyond Vicary's fantasies: The impact of subliminal priming and brand choice. *Journal of experimental social psychology*, *42*(6), 792-798.

Kopetz, C., & Orehek, E. (2015). When the end justifies the means: Self-defeating behaviors as "rational" and "successful" self-regulation. *Current Directions in Psychological Science*, *24*(5), 386-391. https://doi.org/10.1177%2F0963721415589329

Kruglanski, A. W., Chernikova, M., Rosenzweig, E., & Kopetz, C. (2014). On motivational readiness. *Psychological Review, 121*(3), 367–388. https://doi.org/10.1037/a0037013

Locke, E. A., & Latham, G. P. (1990). *A theory of goal setting & task performance*. Prentice-Hall, Inc.

Marien, H., Custers, R., Hassin, R. R., & Aarts, H. (2012). Unconscious goal activation and the hijacking of the executive function. *Journal of personality and social psychology*, *103*(3), 399.

McClelland, D. C., Koestner, R., & Weinberger, J. (1989). How do self-attributed and implicit motives differ? *Psychological Review*, *96*(4), 690.

SENSITIVITY AS IMPLICIT INDICATOR OF GOALS 32McKenna, K. Y., & Bargh, J. A. (1998). Coming out in the age of the Internet: Identity" demarginalization" through virtual group participation. *Journal of personality and social psychology*, *75*(3), 681.

Melnikoff, D. E., & Bailey, A. H. (2018). Preferences for moral vs. immoral traits in others are conditional. *Proceedings of the National Academy of Sciences*, *115*(4), E592-E600.

Neisser, U. (1967). *Cognitive psychology.* New York: Appleton-Century-Crofts.

Nisbett, R. E., & Wilson, T. D. (1977). Telling more than we can know: Verbal reports on mental processes. *Psychological review*, *84*(3), 231.

Nock, M. K., & Banaji, M. R. (2007). Assessment of self-injurious thoughts using a behavioral test. *American Journal of Psychiatry*, *164*(5), 820-823.

Nolan, J. M., Schultz, P. W., Cialdini, R. B., Goldstein, N. J., & Griskevicius, V. (2008). Normative social influence is underdetected. *Personality and social psychology bulletin*, *34*(7), 913-923.

Pessiglione, M., Schmidt, L., Draganski, B., Kalisch, R., Lau, H., Dolan, R. J., & Frith, C. D. (2007). How the brain translates money into force: a neuroimaging study of subliminal motivation. *science*, *316*(5826), 904-906.

Perry, C., & Laurence, J. R. (1984). Mental processing outside of awareness: The contributions of Freud and Janet. *The unconscious reconsidered*, 9-48.

Prout, T. A., Wadkins, M. J., & Kufferath-Lin, T. (2021). *Essential interviewing and counseling skills: An integrated approach to practice*. Springer Publishing Company.

Table 1

*The number of words seen and not and their correspondence with the previously stated or not stated goals*

| Participant # | Number of words seen that correspond to previously stated important goals | Number of words seen that correspond to goals never stated in therapy | Number of words not seen that correspond to previously stated goals | Number of words not seen that correspond to other goals | Number of control words that were seen. | Number of goals they had mentioned in therapy |
|---|---|---|---|---|---|---|
| 1 | 9 | 14 | 2 | 8 | 3 | 2 |
| 2 | 1 | 7 | 3 | 24 | 1 | 1 |
| 3 | 1 | 0 | 0 | 0 | 1 | 1 |
| 4 | 4 | 11 | 3 | 17 | 0 | 2 |
| 5 | 5 | 27 | 2 | 1 | 5 | 2 |
| 6 | 5 | 28 | 2 | 0 | 4 | 2 |
| 7 | 4 | 23 | 0 | 6 | 3 | 2 |